\documentclass[aps,prl,reprint,superscriptaddress,floatfix,nobibnotes]{revtex4-2}

\usepackage{graphicx}
\usepackage{amsmath}
\usepackage{amssymb}
\usepackage{bm}
\usepackage[hidelinks]{hyperref}

\begin{document}

\title{Increased Crystal Nucleation in the Glassy State of a Polymer/Water Mixture}

\author{Mika Kobayashi}
\email{Contact author: kobayashi.phys@gmail.com}
\affiliation{Institute for Materials Chemistry and Engineering (IMCE), Kyushu University, Fukuoka 819-0395, Japan}

\author{Shingo Kobayashi}
\affiliation{Institute for Materials Chemistry and Engineering (IMCE), Kyushu University, Fukuoka 819-0395, Japan}
\affiliation{Graduate School of Biomedical Sciences, Tokushima University, Tokushima 770-8503, Japan}

\author{Kazuyoshi Shiga}
\affiliation{Sumitomo Rubber Industries, Ltd., Hyogo 651-0071, Japan}

\author{Hiroaki Yamada}
\affiliation{Sumitomo Rubber Industries, Ltd., Hyogo 651-0071, Japan}

\author{Masaru Tanaka}
\email{Contact author: tanaka.masaru.763@m.kyushu-u.ac.jp}
\affiliation{Institute for Materials Chemistry and Engineering (IMCE), Kyushu University, Fukuoka 819-0395, Japan}

\date{\today}

\begin{abstract}
Within classical nucleation theory, crystal nucleation requires sufficient supersaturation, but is suppressed at low temperatures due to reduced molecular mobility, leading to a maximum nucleation rate typically above the glass transition temperature $T_g$. Here, we experimentally show that this conventional picture breaks down in a biocompatible polymer poly(2-methoxyethyl acrylate)(PMEA)/water mixture. 
We find that the number of nuclei increases dramatically below $T_g$, with the effective nucleation rate exhibiting an unusual maximum at approximately $T_g-10$ K. Calorimetric experiments reveal a strong temperature-history dependence of cold crystallization 
of water in hydrated PMEA, which is explained by the observed nucleation behavior. We propose that the unusual enhancement of nucleation below $T_g$ provides a key mechanism linking high glass-forming ability and cold crystallization upon heating, thereby explaining the striking asymmetry in crystallization behavior between cooling and heating. 
Our findings suggest that a hidden nucleation-rate maximum below $T_g$ may also underlie cold crystallization in other glass-forming liquids.
\end{abstract}

\maketitle

\newpage
Cold crystallization is widely observed in glass-forming liquids, not only in polymers but also in low molecular liquids and aqueous solutions
\cite{stroble,Wunderlich.1958vwm,Lorenzo.1999,Lu.2000,Tanaka.2000,Ramos.2004,Hatakeyma.2007tc4,
Inaba.2007,Baran.2010,Santis.2011,Kobayashi.2011,Kawakami.2025}. 
It has been used to increase crystallinity and thereby improve heat resistance and mechanical properties of molded polymer products 
\cite{Simmons.2019,Girard.2021,Chivers.1994}. 
The amount of cold crystallization has been practically used as an empirical indicator for evaluating the biocompatible polymers for medical applications
\cite{Tanaka.2000,Tanaka.2000ud6,Sonoda.2020}. 
On the other hand, the microscopic origin of cold crystallization has not been fully understood. 

A liquid can avoid crystallization during cooling, enter a glassy state, and nevertheless 
crystallize relatively easily when reheated through the glass transition. 
Avoiding crystallization on heating is generally much more difficult compared to on cooling. 
This asymmetry is often interpreted in terms of the recovery of molecular mobility during heating, 
especially for the case where the cold crystallization occurs just after the glass transition
\cite{Tanaka.2000,Hatakeyma.2007tc4,Stolte.2013,Klonos.2018,Solarz.2023,Bai.2024,Kostalova.2026}. 
That explanation is incomplete, however, because the same temperature range is also experienced during cooling. A physical description of cold crystallization should therefore focus on when the nuclei are formed, besides subsequent growth.

Classical nucleation theory (CNT) expresses the nucleation rate as
\begin{equation}
J(T)=J_0(T)\exp\left[-\frac{\Delta G^{*}(T)}{k_{\mathrm{B}}T}\right],
\label{eq:J}
\end{equation}
where $J_0$ is a kinetic prefactor and $\Delta G^{*}$ is the free-energy barrier for forming a critical nucleus. Supercooling lowers $\Delta G^{*}$, whereas decreasing temperature reduces molecular mobility and thus $J_0$. The competition between these effects normally produces a maximum in nucleation or growth rate in the supercooled liquid, which has been observed usually above or near the glass transition temperature $T_g$ \cite{Lorenzo.1999, Baran.2010, Huang.2018, Xia.2021,Konishi.2024}. 
The maximum in the nucleation rate generally occurs at a lower temperature than that in the crystal growth rate. Since nucleation is the initial step of crystallization, the temperature range in which nucleation occurs should be 
important for explaining the asymmetry mentioned above. 

Although the mobility decreases, crystal nucleation itself is not prohibited even below $T_g$ 
\cite{Fokin.2003,Fokin.2006,Konishi.2007, Xia.2019, Fokin.2021, Abyzov.2023}. 
There have been many investigations on the effect of thermal history in the glassy state on cold crystallization, 
where the kinetic effect of the relaxation process due to aging has been discussed
\cite{Lu.2000, Santis.2011,Stolte.2013, Fokin.2021, Abyzov.2023}. 
Tammann’s two-stage method has been commonly used in those 
studies below $T_g$
\cite{Androsch.20134a,Androsch.2013,Androsch.2015,Abyzov.2023}
. In this method, nuclei formed at a low nucleation temperature are developed at a higher growth temperature into observable crystals. The nucleation kinetics are analyzed through the number of nuclei directly counted or estimated by thermal analysis. However, the method requires waiting time dependence at the nucleation temperature, which means that the result is influenced by crystal growth and/or physical aging, and it is difficult to separate the effect of nucleation from growth completely. 
In addition, especially in crystalline polymers, the growth rate is generally very high, and it needs a high heating rate to achieve a condition where the contribution of growth rate is neglected. 

Here, we study the relationship between nucleation behavior and cold crystallization 
of a polymer/water mixture, poly(2-methoxyethyl acrylate) (PMEA) and water, 
focusing on the thermodynamic effect of cooling below $T_g$ 
under conditions where the effects of crystal growth and aging can be neglected.
We demonstrate that 
the effective nucleation rate unusually reaches its maximum below $T_g$. 
This behavior well explains the effect of thermal history on the cold crystallization observed in differential scanning calorimetry (DSC). Based on those experimental findings, we propose a scenario for systems that exhibit cold crystallization, where glassy-state nucleation dominates crystallization behavior, including glass-forming ability. 

\begin{figure}[t]
\centering
\includegraphics[width=0.8\columnwidth]{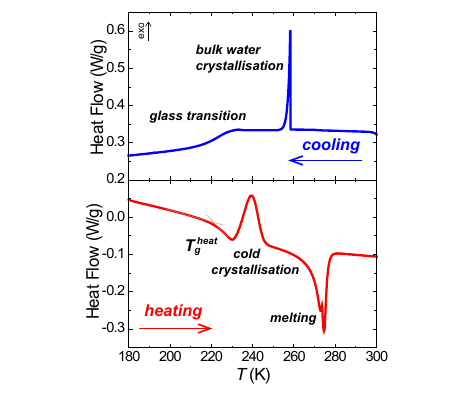}
\caption{DSC cooling and heating curves of water-saturated PMEA. 
The DSC scanning rate is 5 K/min. }
\label{fig:fig1}
\end{figure}

PMEA is an atactic, noncrystalline polymer, and 
the cold crystallization of water in hydrated PMEA has been studied in relation to the hydration behavior of water, 
including intermediate water \cite{Tanaka.2000,Tanaka.2000ud6}. 
The water-saturated PMEA contains $10 \pm 2$ wt\% water and shows a glass transition at $T_g=220~\mathrm{K}$ for a heating rate of 5 K/min. 
This mixture vitrifies even at a slow cooling rate of less than 1 K/min, making it possible to probe the nucleation history with conventional differential scanning calorimetry (DSC) and optical microscopy, without ultrafast temperature control. The system is therefore an appropriate model for separating nucleation from subsequent crystal growth. 
Typical DSC curves on the water-saturated PMEA are shown in Fig. \ref{fig:fig1}. When the sample was cooled, the excess bulk water crystallized around 260 K, showing an exothermic peak, and then the rest of the sample vitrified. 
On heating, before completing the liquid-glass transition, 
an exothermic heat was observed around 240 K, which is the so-called cold crystallization. 
The crystalline phase has been assigned to the hexagonal ice Ih \cite{Kishi.2009,Gemmei-Ide.2026}. 
Then, the melting of the crystal was observed around 270 K. 

\begin{figure}[t]
\centering
\includegraphics[width=0.9\columnwidth]{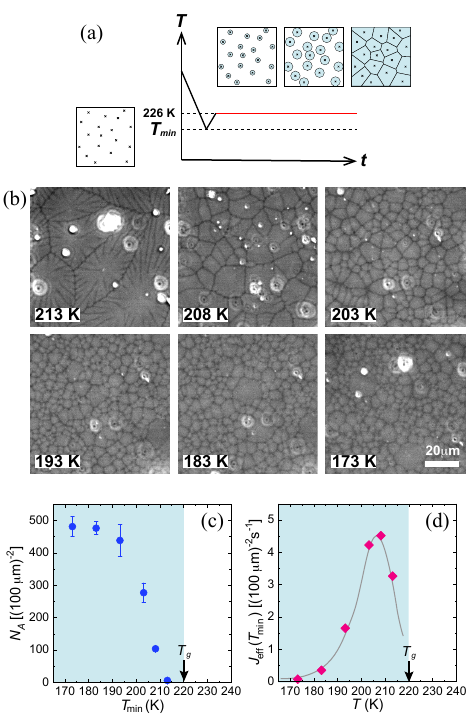}
\caption{(a) Schematic of the protocol used for microscopy. 
(b) Final phase-contrast microscopy images after growth at $226~\mathrm{K}$. 
Numbers at the bottom of each image are the lowest temperature $T_{\min}$ that the system experienced. 
(c) Number of nuclei per unit area  $N_A$ as a function of $T_{\min}$. 
Error bars represent standard deviations across different regions of the sample. 
The glass transition temperature $T_g$ is indicated by an arrow. 
(d) Temperature dependence of effective nucleation rate $J_\mathrm {eff}(T_{\min})$. The solid curve represents a visual guide. 
}
\label{fig:fig2}
\end{figure}

In order to investigate the temperature history effect on cold crystallization, we performed a simple microscopy experiment of the lowest temperature $T_{\min}$-dependence of crystal nucleation.  
The protocol is schematically shown in Fig. \ref{fig:fig2}(a). 
After the rapid cooling at 20 K/min to $T_{\min}$, the sample was immediately heated to 226 K $>$$T_g$ and kept at the temperature to let crystal growth complete. We chose the same growth temperature 
at 226 K for different values of $T_{\min}$. 
The holding time at $T_{\min}$ was less than 1 s. 
The kinetics are sufficiently slow for the system and thus, 
under this protocol, the temperature dependence of the growth rate or the effect of aging can be neglected. 

The microscopy images when the observed crystal growth completed at the grain boundary are shown in Fig. \ref{fig:fig2}(b) for different values of $T_{\min}$. Note that we observed crystal growth close to the glass substrate, indicating that the crystallization would be driven by heterogeneous nucleation. 
When the sample was cooled only to $213~\mathrm{K}$, a small number of grains grew to large sizes. Decreasing $T_{\min}$ produced many more grains, and the grain size correspondingly decreased. 

From those images, we counted the number of crystal grains and estimated the number of crystal nuclei per unit area, $N_A$, 
which represents the number density of nuclei surviving the thermal cycle. 
The result is plotted in Fig. \ref{fig:fig2}(c). 
It is very clear that $N_A$ dynamically increased below $T_g$. 
The additional nucleation during the isothermal growth step at $226~\mathrm{K}$ was negligible on the time scale of the experiment, and the measured growth rate did not show a corresponding dependence on $T_{\min}$, 
indicating that we may rule out the contribution of the crystal growth rate to $N_A$. 

We estimate the temperature dependence of the nucleation rate from $N_A$. 
The number density $N_A(T_{\min})$ is a value accumulated 
while traversing the temperature range down to  $T_{\min}$ and during
the subsequent heating. 
 Because the sample was immediately reheated after reaching $T_{\min}$, 
we assume nucleation during an isothermal hold can be neglected. 
Within a local-rate description, and apart from contributions common to all
thermal cycles, $N_A(T_{\min})$ can therefore be expressed as
\begin{equation}
N_A(T_{\min}) =
N_0+
\int_{T_{\min}}^{226\,{\mathrm K}}
\left[
\frac{J_{\mathrm c}(T)}{\beta_{\mathrm c}}
+
\frac{J_{\mathrm h}(T)}{\beta_{\mathrm h}}
\right] dT ,
\end{equation}
where $N_0$ denotes contributions independent of $T_{\min}$, 
$J_{\mathrm c}$ and $J_{\mathrm h}$ are the areal nucleation rates during
cooling and heating, respectively, and 
$\beta_{\mathrm c} > 0$ and $\beta_{\mathrm h} > 0$ are the cooling and heating rate, respectively. 
Differentiation with respect to $T_{\min}$ extracts the local
contribution at the lowest temperature reached,
\begin{equation}
-\frac{dN_A}{dT_{\min}}
=
\frac{J_{\mathrm c}(T_{\min})}{\beta_{\mathrm c}}
+
\frac{J_{\mathrm h}(T_{\min})}{\beta_{\mathrm h}} .
\end{equation}
Thus, although $N_A$ is an accumulated quantity over the thermal trajectory, its derivative with respect to $T_{\min}$ probes nucleation in the vicinity of $T_{\min}$. In the present experiment, $\beta_c=\beta_h=\beta=20$ K/min. We define the effective nucleation rate $J_\mathrm {eff}(T_{\min})$ as the average of the nucleation rates during cooling and heating at $T_{\min}$:

\begin{equation}
J_\mathrm {eff}(T_{\min})
\equiv \frac{J_c(T_{\min})+J_h(T_{\min})}{2}
=-\frac{\beta}{2}\frac{dN_A(T_{\min})}{dT_{\min}}.
\label{eq:Jeff}
\end{equation}
Here, $J_\mathrm {eff}$ is an effective nucleation rate per unit area, evaluated under continuous cooling and heating, and is not necessarily identical to the steady-state nucleation rate at a constant temperature, as nonstationary effects may occur during temperature change  \cite{Shneidman.1998}. 
Applying this result to Fig. \ref{fig:fig2}(c), 
we obtain $J_\mathrm {eff}(T_{\min})$ in Fig. \ref{fig:fig2}(d). 
Remarkably, $J_\mathrm {eff}(T_{\min})$ reaches its maximum below $T_g$. 
The temperature of the nucleation rate maximum, $T_{\max}$, is around $ T_g- 10$ K. 
This result indicates that 
the lowest temperature the system experienced dominates the increase in $N_A$ rather than 
the elapsed time for the temperature change. 

\begin{figure}[t]
\centering
\includegraphics[]{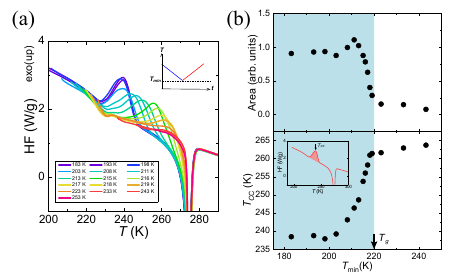}
\caption{Lowest temperature $T_{\min}$-dependence of cold crystallization measured by DSC. (a) Heating curves at 5 K/min after cooling to different $T_{\min}$ values. 
The data are shifted to 0 W/g at 323 K. 
The legend gives corresponding colors for $T_{\min}$. 
The DSC measurement protocol is given in the inset schematic, where the cooling/heating rate was 5 K/min. 
(b) (top)Amount of cold crystallization as a function of $T_{\min}$, 
estimated from the exothermic peak of cold crystallization shown in (a). 
The definition of the integration and peak temperature is given in the inset schematic. 
(bottom)Peak temperature of the cold crystallization peak. 
}
\label{fig:fig3}
\end{figure}

Next, we show the effect of thermal history on the DSC curves. 
Figure \ref{fig:fig3}(a) shows the DSC heating curves measured after cooling down to different lowest temperatures. 
In this experiment, the cooling and heating rate was 5 K/min. 
The data show the strong effect of thermal history, as the heating curves systematically change with changing $T_{\min}$. For the data cooled to the highest temperature, 253 K, there is almost no obvious crystallization on heating, and a melting peak is regarded as ice crystals of bulk water crystallized on cooling. 
With a decrease in $T_{\min}$, the peak appeared, became larger, and shifted to a lower temperature. The shift of the peak stopped when $T_{\min}$ was below 198 K. Note that a small endothermic peak around 230 K observed for the data of 223 K is regarded as due to an enthalpy relaxation on the glass transition. 

The amount of crystallization was estimated by integrating the peak area of the cold crystallization and plotted against $T_{\min}$ in Fig. \ref{fig:fig3}(b). The crystallization behavior clearly changed around a boundary temperature near $T_g$. 
When $T_{\min}$ was above $T_g$, the amount of crystallization was very small. It increased dramatically when $T_{\min}$ was lowered below $T_g$. The value seemed to saturate to a constant value at lower $T_{\min}$. 
This is consistent with the increase in the number of crystal nuclei shown in Fig. \ref{fig:fig2}(c). 
The peak temperature $T_\mathrm {CC}$ is shown in Fig. \ref{fig:fig3}(b) 
as a function of $T_{\min}$. 
 It started to shift towards lower temperatures with a decrease in $T_{\min}$, 
indicating that there is a boundary temperature around $T_g$. 

The properties observed in microscopy and thermal measurements 
are consistent with each other. 
They started to change around similar boundary temperatures close to $T_g$. 
Our results can be explained as follows. 
When the system is not cooled sufficiently (high $T_{\min}$), $N_A$ is small and 
the system has to complete the growth of crystal grains to the boundary of other grains or wait for additional nucleation. 
Thus, it takes time to complete the cold crystallization, resulting in the transition peak becoming broad and $T_\mathrm {CC}$ becoming high. When $N_A$ increases after the system experienced $T_{\min}$ below $T_g$, many nuclei are already generated, which means fast completion of crystallization for the same growth rate. At lower $T_{\min}$, $N_A$ becomes saturated and almost the same. Accordingly, the crystallization peaks on heating become almost identical.  
All the behavior seems to originate from the unusual nucleation rate maximum below $T_g$. 
This provides a kinetic explanation for the cold crystallization behavior previously observed in hydrated PMEA 
\cite{Tanaka.2000}. 

\begin{figure}[t]
\centering
\includegraphics[width=0.7\columnwidth]{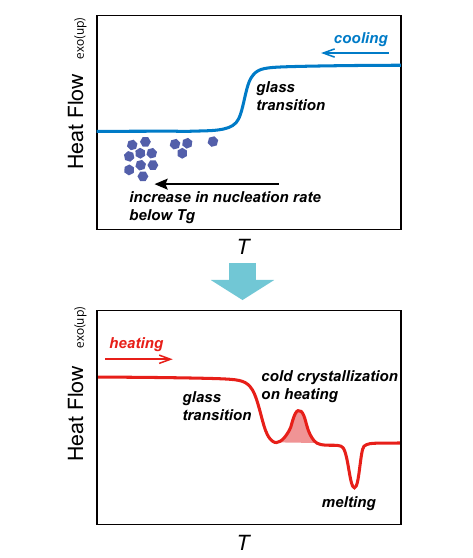}
\caption {Schematic illustration of the mechanism underlying the relationship between 
glass-forming ability and cold crystallization, which is reflected in DSC curves.  
}
\label{fig:fig4}
\end{figure}

Based on the above experimental results, 
we suggest the following fundamental scenario to explain the mechanism of cold crystallization, 
and the asymmetry in the crystallizability between cooling and heating.  
The schematic illustration to explain the behavior observed in DSC is shown in Fig. \ref{fig:fig4}.  When the system has the property that the nucleation rate increases below $T_g$, 
due to the low nucleation rate at $T>T_g$ , the system has high glass-forming ability and is able to vitrify on cooling even at a slow cooling rate. 
Below $T_g$, the crystallization is not detectable in conventional DSC because of the low mobility. 
On the other hand, on heating from temperatures below $T_g$, the system has already experienced a temperature range where the nucleation rate increases, meaning that there are already many crystal nuclei produced there.  
The crystal grows when the mobility increases at $T_g$, so crystallization on heating is inevitable. 

When the nucleation rate maximum temperature $T_{\max}$ is above $T_g$, 
as commonly observed, 
the system would easily crystallize on cooling, indicating that the system needs a high cooling rate to vitrify. When the system is heated before reaching the temperature range of high $J$, crystallization would not be observed. This means that whether a supercooled liquid exhibits cold crystallization depends on 
the temperature range into which the system is cooled.
The property that the temperature range of high $J$ is below $T_g$ would cause the glass-forming ability and the asymmetry in the ease of crystallization between cooling and heating. 

The remaining question is under what conditions the nucleation rate can reach its maximum below $T_g$. 
We address this question by considering two competing factors in Eq. \ref{eq:J}. 
For homogeneous nucleation, increasing supercooling reduces the thermodynamic barrier $\Delta G_\mathrm {hom}^*$, whereas decreasing molecular mobility suppresses the kinetic prefactor \(J_0\). The balance between these two contributions determines the temperature $T_{\max}^\mathrm {hom}$ of maximum nucleation rate. 
Using the fragility index $m$ of glass-forming liquids
\cite{Boehmer.1993}, 
we derive a condition for $T_{\max}^\mathrm {hom}<T_g$ from the competition between the kinetic prefactor and the thermodynamic nucleation barrier in Eq. \ref{eq:J} (see End Matter):
\begin{equation}
\frac{m\ln 10}{T_g}
<
\left.
\frac{d}{dT}
\left(
\frac{\Delta G_\mathrm {hom}^*}{k_BT}
\right)
\right|_{T_g}.
\end{equation}
This condition indicates that stronger liquids with small $m$ would have a possibility to satisfy it 
\cite{James.1985, Gupta.2016tz}.
Consistently, theoretical analyses have shown that $T_{\max}$ shifts to higher temperatures with increasing fragility and to lower temperatures with increasing thermodynamic nucleation barrier 
\cite{Schmelzer.2015}. 
On the other hand, even in the experimentally studied strongest glass formers, such as silicate glass formers, 
the maximum homogeneous nucleation rate occurs at or above $T_g$ 
in most cases 
\cite{James.1985, Fokin.2003, Gupta.2016tz}. 
Indications that $T_{\max}^\mathrm {hom}$ lies below $T_g$ have rarely been reported 
\cite{Wakasugi.1999, Deubener.2000}. 
For the present polymer/water mixture, which is expected to behave as a usual molecular or polymeric glass former, satisfying the condition may require an additional mechanism, such as a fragile-to-strong transition 
\cite{Cerveny.2016h0s,Shi.2018} or a pronounced reduction in the nucleation barrier associated with the glass transition. 

Further, the present maximum occurs below $T_g$ even though the observed nucleation is predominantly heterogeneous. Within the simplest CNT description, heterogeneous barrier lowering shifts $T_{\max}$ to 
a higher temperature relative to homogeneous nucleation 
\cite{Gupta.2016tz}. 
Thus, the underlying homogeneous-nucleation maximum would be expected to lie at an even lower temperature, $T_{\max}^\mathrm {hom}<T_{\max}^\mathrm {het}<T_g$, emphasizing the unusually strong tendency for nucleation in the glassy regime.

In summary, 
we experimentally established a mechanism linking crystal nucleation, 
glass-forming ability, and cold crystallization upon heating. 
This mechanism explains the observed behavior in the present system, and the underlying idea is not system-specific but can be applied to other systems: a nucleation-rate maximum below $T_g$ can account for both high glass-forming ability and cold crystallization upon heating. This suggests that a hidden nucleation-rate maximum below $T_g$ may exist in other systems, especially those exhibiting cold crystallization close to $T_g$.
Our analysis suggests the unusual behavior originates from 
fragility or reduction of the nucleation barrier.
The behavior is unusual for the typical value of the fragility index for a polymer/water mixture, and 
further investigation is needed. 
Fragile-to-strong transition in water might be a possibility related to the behavior 
\cite{Cerveny.2016h0s,Shi.2018}.
Alternatively, a strong reduction in the nucleation barrier, possibly associated with dehydration, phase separation, or a liquid-liquid phase transition, may play a role. 
Our results provide a new perspective on nucleation in glassy systems and may also offer insight into applications including biocompatible polymers, heat-storage materials, and polymer processing technology.

\begin{acknowledgments}
{\it Acknowledgments--}
The authors thank Hajime Tanaka, Koji Fukao, and Takahisa Anada for valuable discussions 
and Yukiko Tanaka for technical support. 
This work was supported in part by 
JSPS KAKENHI (Grant Nos. JP22H00591, JP24K22382, and JP26K24019) 
from the Japan Society for the Promotion of Science; 
Dynamic Alliance for Open Innovation Bridging Human, Environment, and Materials; 
Integrated Research Consortium on Chemical Sciences; 
and by Sumitomo Rubber Industries, Ltd. 
\end{acknowledgments}

\paragraph*{Data availability--}
The data supporting the findings of this study are available from the authors upon reasonable request. 

\bibliography{PMEA1}

\section{End Matter}

\paragraph*{Materials.}
We used laboratory-synthesised PMEA (weight average molecular mass $M_w=8.9 \times 10^4$, number average molecular mass $M_n=2.8 \times 10^4$, $M_w/M_n=3.2$, as determined by GPC using THF as the eluent and calibrated with polystyrene standards) for the sample. Water-saturated PMEA was prepared and used for experiments. 
PMEA (poly(2-methoxyethyl acrylate)) was synthesized by conventional free-radical polymerization according to a previously reported procedure\cite{Sato.2015}. 
After the synthesis, the polymer was stored in a refrigerator. Before the preparation of water-saturated PMEA, it was dried again at 363 K for 8 h in a vacuum to remove residual solvents. Then, pure water was added to the dried sample in a glass tube, and the polymer was kept at room temperature to allow it to absorb water. Note that PMEA of molecular mass in this study is poorly soluble in water. The typical amount of polymer for preparation of water-saturated PMEA was a few mL, and the amount of water was more than two times the volume of polymer. For this amount, the water content is equilibrated after the sample has been kept for more than two weeks. The equilibrated water content $w_{\mathrm water}$ was 10 $\pm$ 2 wt\%, estimated from the mass difference before and after drying, $m_{\mathrm wet}$ and $m_{\mathrm dry}$; $w_{\mathrm water} = (m_{\mathrm wet}-m_{\mathrm dry})/m_{\mathrm wet}$. 
The glass transition temperature $T_g^{heat}$ of the water-saturated PMEA was 220 K as an onset temperature for a heating rate 
of 5 K/min. 

\paragraph*{DSC measurements.}
The DSC measurements were performed using a calorimeter (SII NanoTechnology Inc. X-DSC7000). The sample weight was 5-10 mg. The cooling and heating rates were both 5 K/min. We used standard aluminum sample pans and used air as the reference. All the measurements were performed under N$_2$ atmosphere. 

\paragraph*{Microscopy.}
We performed phase contrast microscopy using a microscope (Olympus BX53, objective lens 20X, N.A. 0.45). The images were taken with a scientific CMOS camera (Andor Zyla) controlled by software (Nikon NIS-Elements). The sample was sandwiched between two glass plates without a spacer. The water containing PMEA was extracted with toothpicks and placed on a glass plate. A cover glass was placed on it, and the glass plates were pressed by hand to make it thinner and sealed with epoxy adhesive. The sample thickness was expected to be 10 to 20 $\mu$m. Temperature was controlled by a temperature stage (Linkam LTS420) with liquid nitrogen with an accuracy of ± 0.1 K. 

\paragraph*{Condition for the maximum nucleation rate below $T_g$.}
Starting from Eq. (1), the temperature dependence of the homogeneous nucleation rate is given by 
\begin{equation}
\frac{d\ln J_\mathrm {hom}}{dT}
=
\frac{d\ln J_0}{dT}
-
\frac{d}{dT}
\left(
\frac{\Delta G_\mathrm {hom}^*}{k_BT}
\right).
\end{equation}
The first term represents the kinetic contribution associated with molecular mobility, whereas the second represents the thermodynamic enhancement of nucleation arising from the reduction of the nucleation barrier upon cooling. 
Assuming that the nucleation rate eventually decreases at sufficiently low temperatures due to reduced molecular mobility, a maximum below $T_g$ requires that $J_\mathrm {hom}$ is still increasing upon cooling at $T_g$, or equivalently
\begin{equation}
\left.
\frac{d\ln J_\mathrm {hom}}{dT}
\right|_{T_g}<0.
\end{equation}
If the dominant temperature dependence of the kinetic prefactor follows the mobility, $J_0\propto D\propto\tau^{-1}$, 
where $D$ and $\tau$ are the molecular diffusion coefficient and structural relaxation time, respectively, 
the conventional fragility 
\cite{Boehmer.1993}
\begin{equation}
m=
\left.
\frac{d\log_{10}\tau}
{d(T_g/T)}
\right|_{T_g}
\end{equation}
gives, without assuming a specific Vogel-Fulcher-Tammann form,
\begin{equation}
\left.
\frac{d\ln J_0}{dT}
\right|_{T_g}
\simeq
\frac{m\ln10}{T_g}.
\end{equation}
Thus, the condition for 
maximum nucleation rate temperature $T_{\max} < T_g$
becomes
\begin{equation}
\frac{m\ln 10}{T_g}
-
\left.
\frac{d}{dT}
\left(
\frac{\Delta G_\mathrm {hom}^*}{k_BT}
\right)
\right|_{T_g} < 0.
\end{equation}
This condition is achieved by a relatively weak kinetic slowdown and/or by a strong temperature dependence of the thermodynamic nucleation barrier.
This result makes explicit that $T_{\max}^\mathrm {hom}<T_g$ is not prohibited by CNT, although its occurrence depends on the competition between the kinetic and thermodynamic contributions. Previous CNT analyses similarly found that $T_{\max}$ increases with increasing fragility and decreases with increasing thermodynamic nucleation barrier 
\cite{Gupta.2016tz}. 

The nuclei observed here form preferentially near the glass substrate, indicating predominantly heterogeneous nucleation. 
Heterogeneous nucleation reduces the nucleation barrier, and the maximum nucleation rate 
is expected to occur at a higher temperature than that for homogeneous nucleation. 
Therefore, the corresponding maximum for homogeneous nucleation would be expected to occur at an even lower temperature than the experimentally observed maximum: 
\begin{equation}
T_{\max}^\mathrm {hom}
<
T_{\max}^\mathrm {het}
<
T_g.
\end{equation}
This criterion depends only on the relative temperature dependences of the kinetic prefactor and the nucleation barrier as $T_g$ is approached, and does not require an extrapolation of a VFT form below $T_g$.

\end{document}